\documentstyle[12pt]{article}
\begin{document}
\begin{flushright}
CERN-TH/96-333
\end{flushright}
\vspace{0.5cm}

\begin{center}
{\bf RESONANT SCATTERING OF ULTRAHIGH-ENERGY NEUTRINOS\\
IN THE  $SU(3)_C \otimes SU(3)_L \otimes  U(1)_N$ MODEL }\\
\vspace{2cm}
{\bf  Hoang Ngoc Long\footnote{Permanent address: 
Institute of Theoretical
Physics, National Centre for Natural Science and Technology, 
P. O. Box 429, Bo Ho, Hanoi 10000, Vietnam}}\\
{\it  Theory Division, CERN,
CH - 1211 Geneva 23,
Switzerland}\\
\vspace{2cm}

{\bf Abstract}\\
\end{center}
The lepton  family-number changing $\Delta L_{(l)} = 2$ 
neutrino reactions in the framework of the $SU(3)_C 
\otimes SU(3)_L \otimes  U(1)_N$ model with
right-handed neutrinos are presented. 
These processes have a much-enhanced cross section as a result
of $s$-channel resonant bilepton $Y^\pm$ production and they should
be added to ultrahigh-energy neutrino interactions.   

\vspace{1cm}
PACS number(s): 13.15.+g, 14.60 St, 14.70 Pw, 95.55. Vj.

\vspace{3.7cm}

\begin{flushleft}
CERN-TH/96-333\\

November 1996
\end{flushleft}
\newpage
\noindent
\hspace*{0.5cm}Recently physicists have  been intrigued by 
ultrahigh-energy (UHE) cosmic-ray neutrinos, since there
is evidence that Cygnus X-3 is likely to be a potent source. 
 With such a high energy $\sim 6 \times  10^{6}$  GeV, 
effects of new physics will certainly play  a role. 
By common estimates~\cite{lang}, many versions of extended gauge
theories predict a mass scale for new physics  in the range 
of few TeV. The   recently proposed  models based on the  
$SU(3)_C \otimes SU(3)_L \otimes  U(1)_N$ (3 -- 3 -- 1) gauge
group~\cite{svs,ppf,flt} have the nice feature that, 
in addition to the QCD asymptotic freedom, they predict 
that the number of families is to be equal to 3. 
Four new charged gauge bosons responsible for the violation of 
the lepton  family-number  (bileptons) have masses in the 
limit~\cite{dng}: 270\  GeV $\le m_Y \le$ 550\ GeV, and
the limit for the mass of the new heavy neutral gauge boson
 1.3\  TeV $\le m_{Z_2} \le$ 3.1\ TeV. The {\it upper}
limit arises from the model's consequence that $\sin^2
\theta_W(M_{Z'})$ is to be less than 1/4. 
Therefore, it seems appropriate in this context
to consider new effects such as fermion number changing 
processes to the known ones in the standard model (SM). In the 
3 -- 3 -- 1 models,
the process with $\Delta L_{(l)} =2$: $e^- \nu_e \rightarrow 
\mu^- \nu_\mu $ can be possible at high energy due to 
the new bileptons $Y^{\pm}$.  In this Note we consider new 
physical processes which have not  been considered
in~\cite{cq}.

Interactions of UHE neutrinos with nucleons have been considered by
Quigg, Reno and Walker~\cite{cq,qrw} in the framework of the SM and 
the renormalization-group-improved parton model. Neutrino-electron 
cross sections are  generally small with respect to 
neutrino-nucleon ones, because of the electron's small mass. 
However, the resonant cross section considered first by 
Glashow~\cite{sg} is an exceptional case. It has a resonant
peak at the laboratory energy of the incoming neutrino $E_{\nu e} 
= m_W^2/2m_e \simeq 6 \times 10^6$ GeV (for $e^-\nu_\mu 
\rightarrow \mu^-\nu_e$),  and is larger than
the $\nu N$ cross-section at any energy up to $10^{12}$ GeV
(see Figs. 10 in~\cite{cq,ghs}).
Besides usual neutrino-nucleon, neutrino-electron scatterings,
in the 3 -- 3 -- 1 model,  the following processes exist:
\begin{eqnarray}
e^-\ \nu_e &\rightarrow& l^-\ \nu_l,  \   l = \mu , \tau,\\
\nu^a\  \nu^a &\rightarrow & \nu^b\  \nu^b,  \   a \neq b,
\ a,\ b = e, \ \mu, \  \tau.
\end{eqnarray}
Process (1) exists in both 3 -- 3 -- 1 versions: the minimal
model~\cite{ppf,fhpp} and the model with right-handed
neutrinos~\cite{svs,flt,prd3}, while
process (2) is specific of the model with right-handed neutrinos.

The necessary interaction Lagrangians for these  processes are given
in Ref.~\cite{prd3}:
\begin{equation}
{\cal L} = - \frac{g}{\sqrt{2}}[\bar{(\nu^c_L)}^a\gamma^\mu 
e^a_LY^+_\mu + \bar{\nu}^a_L\gamma^\mu (\nu^c_L)^aX^0_\mu + 
\mbox{h.c.}],
\end{equation}
where $a$ = 1, 2, 3 is the generation index, and
$g=e/ \sin\theta_W $ ($\theta_W$ is being  the Weinberg angle).

An amplitude for the process is given by:
\begin{eqnarray}
T_{if}=&-&\frac{g^2}{8}\left\{\bar{u}_e({\bf p},\sigma)
(1+\gamma_5) \gamma_\nu v_{(\nu_e)}({\bf q}, \lambda)
\right.\nonumber\\
&\times&\left.D^{\nu \rho}(p + q) \bar{v}_{(\nu_\mu)}
({\bf q}_1, \lambda_1)
\gamma_\rho (1 - \gamma_5) u_{(\mu)}({\bf p}_1,\sigma_1)\right\},
\end{eqnarray}
where  we used $ v({\bf q},\lambda) = C \bar{u}^T({\bf q}, \lambda)$
and $ \bar{v}({\bf q}, \lambda) = u^T({\bf q},\lambda) C^{-1}$.
In (4)  $p,\ q$ and $p_1, \ q_1$ are momenta of leptons and neutrinos
in the initial and final states, respectively,
and for simplicity we take $D^{\mu\nu}(k)= - g^{\mu\nu}/(k^2-m_Y^2 
+ i m_Y \Gamma_Y)$.

After some manipulations we get the cross section in the laboratory
frame, in which the electron is at rest:
\begin{equation}
\sigma(e^- \nu_ e \rightarrow \mu^- \nu_\mu)=\frac{G_F^2 \ m_W^4\ m_e 
\ E_{\nu e}}{3 \pi [ (s-m_Y^2)^2 + m_Y^2 \Gamma_Y^2 ]},
\end{equation}
where $E_{\nu e},..., m_e,...$ are the laboratory energy of incoming 
neutrinos and the mass of particles, and
$s = 2 m_e E_{\nu e}$. Here we neglected the masses of neutrinos and
used
\[\frac{ G_F}{\sqrt{2}} = \frac{g^2}{8 m_W^2}, \]
which is correct  at least at the  tree level.
Note that the processes considered here 
occur in the $s$ channel with intermediate bileptons $Y^\pm$.

From (5), it follows that the cross section is enhanced at 
the resonant energy:
$E_{\nu e}^{res} = \frac{ m^2_Y}{2 m_e} \simeq  160 \times
10^6$  GeV.
Here we take the mass of the $Y$ boson to be 400 GeV, which came
from the muon decay experiments~\cite{rpp}
\[ R = \frac{\Gamma (\mu^- \rightarrow e^- \nu_e \bar{\nu}_\mu)}
{\Gamma (\mu^- \rightarrow e^- \bar{\nu}_e \nu_\mu)}
\sim \left(\frac{m_W}{m_Y}\right)^4 < 1.2\%, \  90\% \  CL,\]
(the lower limit is $m_Y \ge 270$ \ GeV).

Let us make {\it rough} numerical estimates. Assuming that
the exotic quarks are heaver than bileptons, we have then
\[ \Gamma_Y = \sum_{l=e,\mu,\tau}\Gamma(Y^- \rightarrow l\
\nu_l) = \frac{G_F\ m_W^2 \ m_Y}{2\sqrt{2}\ \pi}.\]
Considering the effects of the $Y^\pm$ resonance region, 
$(m_Y^2/2 m_e - 2 \Gamma') \leq E_{\nu e} \leq (m_Y^2/2 m_e
 + 2 \Gamma')$, where $\Gamma' = \Gamma_Y m_Y/2 m_e $ , 
we get  the average cross section $\sigma_{int} \simeq 
8 \times 10^{-33} cm^2$, while the integrated cross section
for $\bar{\nu}_e e \rightarrow \bar{\nu}_\mu \mu$  in the SM
is~\cite{cq}: $5.38 \times 10^{-32} cm^2$.

Similar calculation gives the cross section  in the c. m. frame
for  process (2) as follows:
\begin{equation}
\sigma(\nu_e \nu_e \rightarrow \nu_\mu \nu_\mu)=\frac{G_F^2 \  
m_W^4\  s}{6 \pi\ [  (s-m_X^2)^2 + m_X^2 \Gamma_X^2 ]}.
\end{equation}
Here we have a resonance at $s \simeq m_X^2$ and the mass of $X$ 
is equal to
$m_Y$, due to  the symmetry-breaking hierarchy~\cite{prd3}.
Process (2) has $\Delta L_{(l)} = 2$ and with intermediate neutral 
gauge bosons $X$. It is very similar to neutrino oscillations, 
but here without requiring that the neutrinos have masses.
In (5) and (6) we have considered that at high energy, neutrinos 
have two spin states

We discuss now  the selection problem. Note that in the SM 
and in the 3 -- 3 -- 1 model we have a similar process:
\begin{equation}
e^- \ \nu_\mu \rightarrow \mu^- \ \nu_e. 
\end{equation}
This means that in the final state of two different processes
(1) and (7) we have the same charged lepton $\mu^-$.
To distinguish these different initial states, we have to
account for helicity. In (1)  the participating particles
have different helicities, while in (7) they have the same
helicities.
We conclude, then, future neutrino telescopes should have
new characteristic, namely helicities.

In this Note we have discussed UHE neutrino-electron reactions 
in which there are lepton family-number violations 
($\Delta L_{(l)} = 2$)
and   resonant peak  in the $s$ channel.
The processes considered here will find  applications in
astrophysics and cosmology, and we will return to these 
problems in the future.

Many questions on neutrinos are still unresolved. Moreover, 
neutrinos and especially UHE ones can bring valuable information
directly from their source. As highly stable, weakly interacting
and neutral, neutrinos can go   
through intervening magnetic fields undeflected. 
The feebleness of neutrino interactions  means that neutrinos
can give exciting  information about  cosmology that other radiation 
cannot. Therefore, at present, neutrino astronomy~\cite{tot} is 
considered as one of the most prospective subjects.
The extraterrestrial neutrinos  provide valuable information
about our Universe, and
UHE neutrinos will give a clue about the origin of cosmic
rays, but vast detectors with new characteristics are required. 
Large volume high energy neutrino telescopes with potential
effective areas of the order of $10^9 \ cm^2 $  and volumes 
greater than $2\times 10^{13} \ cm^3$~\cite{nt} are being
planned. We expect that neutrino telescopes
will detect UHE neutrinos from extraterrestrial sources, and will
begin to test models for neutrino production in active galactic 
nuclei, and the results will confirm or rule out our 
model. \\[0.3cm]

\hspace*{0.5cm}The author would like to thank  P. Aurenche, 
G. Belanger, F. Boudjema, J. Kaplan, O. Pene and X. Y. Pham 
for valuable discussions and remarks. He also 
thanks the  CERN Theory Division
for financial support and hospitality. This paper is
dedicated to Professor Abdus Salam - one of the founders
of the electroweak theory - to whom the author is deeply 
indebted. \\[0.5cm]

\end{document}